\documentclass{appolb}
\usepackage{cite,here,epsfig}

\begin{document}
\title{ Experimental study of the $\eta$-meson interaction with two-nucleons 
\thanks{Presented at the international conference MESON 2002}
}
\author{
         P.~Moskal$^{1,2}$,
         H.-H.~Adam$^3$,
         A.~Budzanowski$^4$,
         R.~Czy{\.{z}}ykiewicz$^2$,
         D.~Grzonka$^1$,
         M.~Janusz$^2$,
         L.~Jarczyk$^2$,
         B.~Kamys$^2$,
         A.~Khoukaz$^3$,
         K.~Kilian$^1$,
         P.~Kowina$^{1,6}$,
         N.~Lang$^3$,
         T.~Lister$^3$,
         W.~Oelert$^1$,
         T.~Ro{\.{z}}ek$^{1,6}$,
         R.~Santo$^3$,
         G.~Schepers$^1$,
         T.~Sefzick$^1$,
         M.~Siemaszko$^6$,
         J.~Smyrski$^2$,
         S.~Steltenkamp$^3$,
         A.~Strza{\l}kowski$^2$,
         P.~Winter$^1$,
         M.~Wolke$^1$,
         P.~W{\"u}stner$^5$,
         W.~Zipper$^6$
 \address{ $^1$ IKP, Forschungszentrum J\"{u}lich, D-52425 J\"{u}lich, Germany                \\
           $^2$ Institute of Physics, Jagellonian University, PL-30-059 Cracow, Poland        \\
           $^3$ IKP, Westf\"{a}lische Wilhelms--Universit\"{a}t, D-48149 M\"{u}nster, Germany \\
           $^4$ Institute of Nuclear Physics, PL-31-342 Cracow, Poland                        \\
           $^5$ ZEL,  Forschungszentrum J\"{u}lich, D-52425 J\"{u}lich,  Germany              \\
           $^6$ Institute of Physics, University of Silesia, PL-40-007 Katowice, Poland       \\
         }
}
\maketitle
\begin{abstract}
        By means of the COSY-11 detection system,
 using a stochastically cooled proton beam
 of the {\bf Co}oler {\bf Sy}nchrotron COSY
 and a hydrogen cluster target,
 we have performed a high
 statistics measurement of the $pp \rightarrow pp\eta$ reaction
 at an excess energy of Q~=~15.5~MeV.
 The experiment was based on the four-momentum determination
 of both outgoing protons. This permits to identify $pp\to pp\eta$ events
 and to derive the complete kinematical
 information of the $\eta$pp-system allowing for subsequent
 investigations of the $\eta$p interaction.
 The observed spectrum of the invariant mass of the proton-proton
 system deviates strongly from the phase-space distribution revealing the influence 
 of the final-state-interaction among the outgoing particles or  nontrivial 
 features of the primary production mechanism.
\end{abstract}
  
\section{Introduction}
  Due to the short live time of the  flavour-neutral mesons 
 (eg. $\pi^0,~\eta,~\eta^{\prime},~\omega$),
 the study of their
 interaction with nucleons or other mesons is at present not feasible in direct 
 scattering experiments. One of the methods 
 permitting such investigations is the production of a meson in
 the nucleon--nucleon interaction close to the kinematical threshold 
 or in kinematics regions where the
 outgoing particles possess small relative velocities.
 When 
 the relative kinetic
 energy is in the order of a few MeV, the final state particles remain much
 longer in the range of the strong interaction than the typical life--time of
 $N^{*}$ or $\Delta$ baryon resonances with $10^{-23}\,\mbox{s}$.
 Thus, 
 they can easily experience a mutual interaction before escaping the area of an
 influence of the hadronic force. 
 This interaction modifies the phase-space abundance and changes
 the distributions of the differential cross sections and the magnitude of 
 the total reaction rate. A precise determination of the 
 energy dependence of the total cross section close to the production 
 threshold of the 
 $pp\to pp\eta$~\cite{eta_data}
 and $pn \to d\eta$ reactions~\cite{eta_pn}
 revealed an enhancement 
 at low excess energies generally accepted as
 a signal from the $\eta$--nucleon interaction. A similar effect is also observed
 in the photoproduction of $\eta$ via the $\gamma d \to pn\eta$ reaction~\cite{eta_photo},
 indicating to some extent that the phenomenon is independent of the production 
 process but is rather related to the interaction among
 $\eta$-meson and nucleons in the $S_{11}$ region.
 Interestingly, out of all studied flavour-neutral mesons only the $\eta$-nucleon
 force is strong enough to manifest itself in the 
 excitation function of the total cross section 
 over the overwhelming nucleon-nucleon interaction.
 In the case of the production of other mesons no such enhancement has been observed,
 though the similar experimental precision has been achieved for example for 
 the threshold production of $\pi^{0}$~\cite{pi0_data}
 or $\eta^{\prime}$~\cite{etap_data} mesons.
 Hence, with the up--to--date experimental accuracy, from all $Meson\,NN$--systems the
 $\eta NN$ one reveals by far the most interesting features.  
\vspace{-0.5cm}
\begin{figure}[H]
\parbox{0.5\textwidth}{\centerline{
        \epsfig{figure=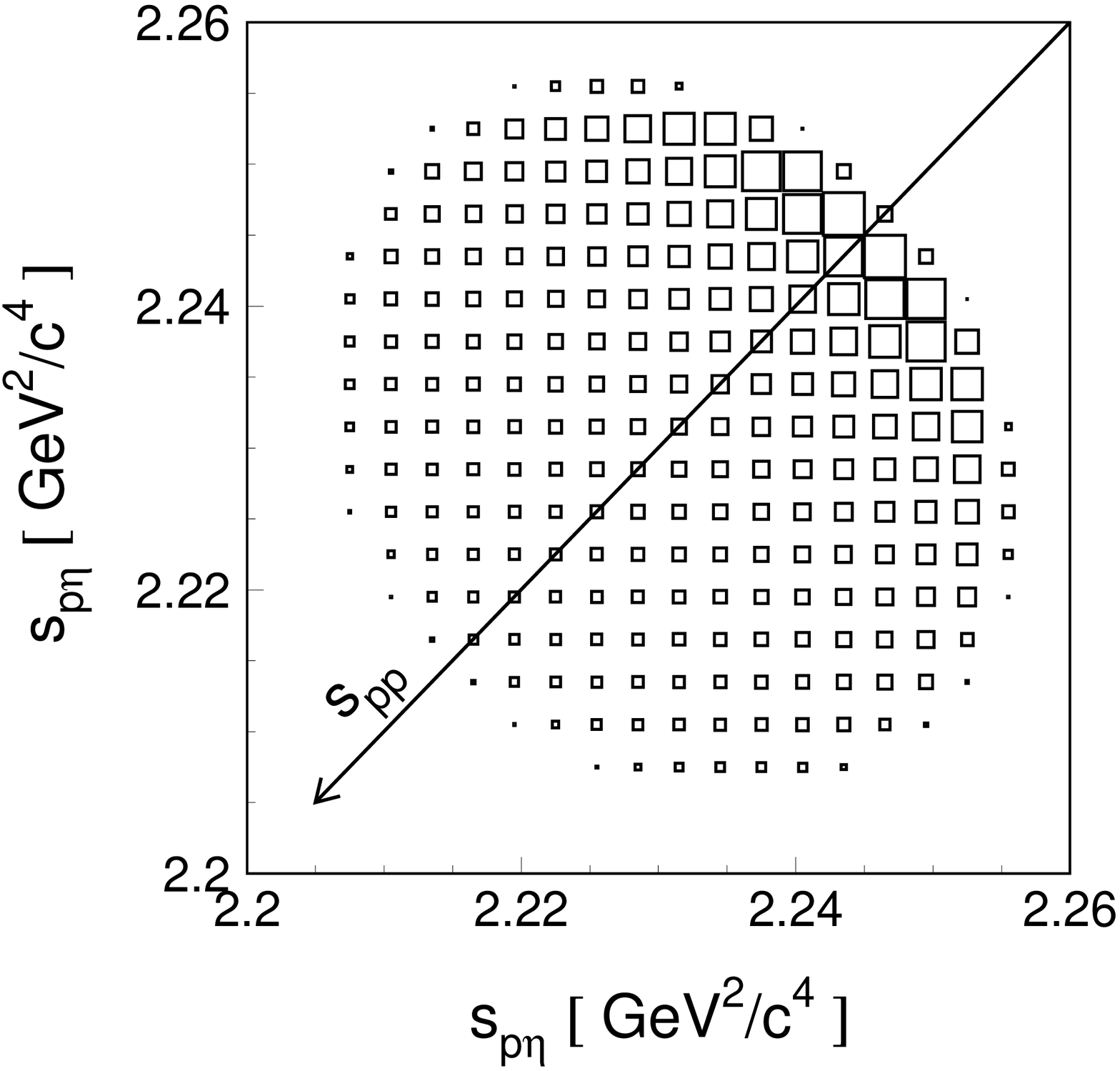,width=0.4\textwidth,angle=0}}
}
\parbox{0.5\textwidth}{\centerline{
        \epsfig{figure=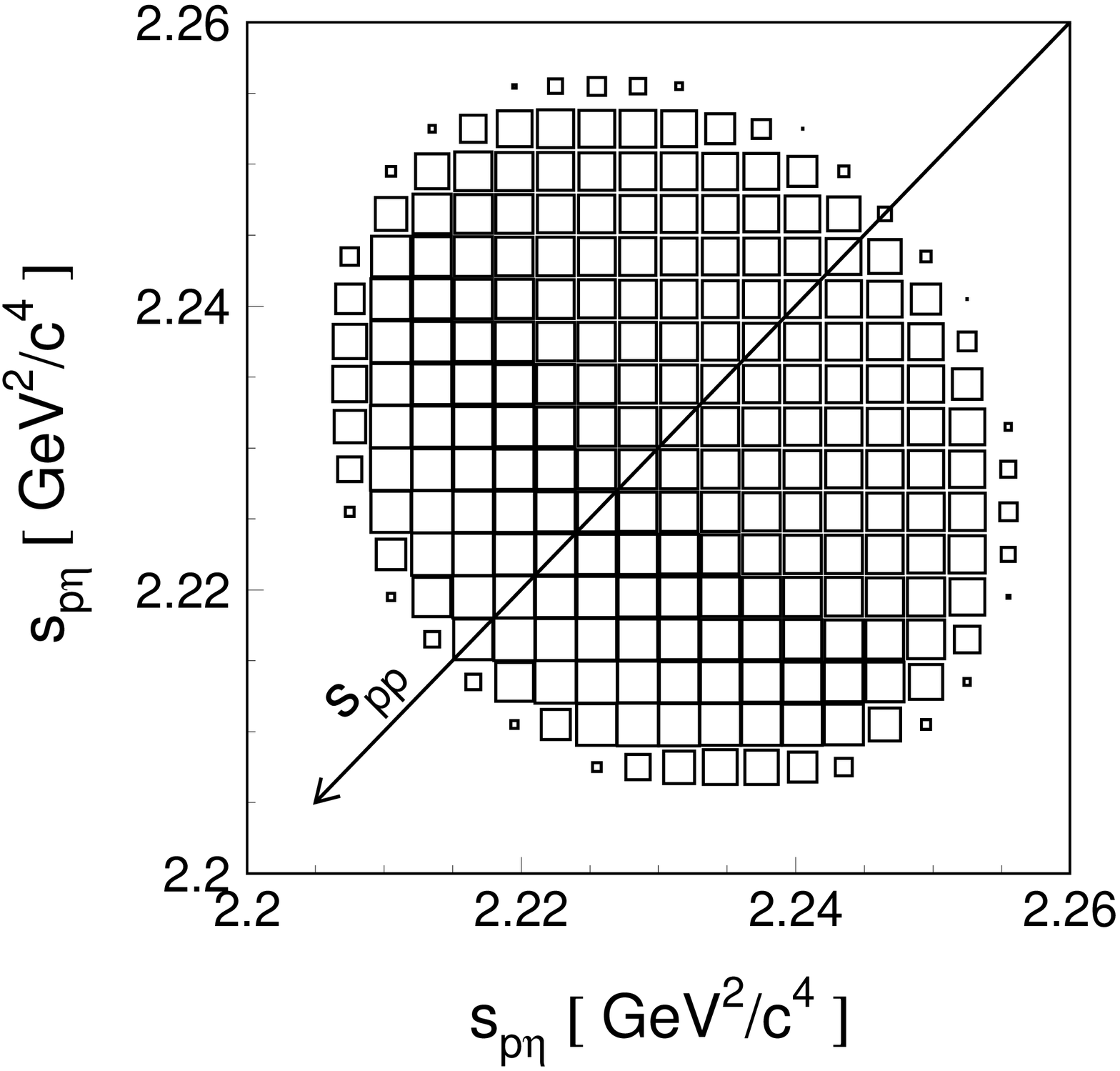,width=0.4\textwidth,angle=0}}
}
\parbox{1.0\textwidth}{
  \caption{ \label{dalitz_mc}
     Monte-Carlo simulations for the $pp \to pp\eta$ reaction at Q~=~16~MeV:
     (left) Phase-space density distribution modified by the proton~-~proton final state interaction. 
     (right) Phase-space density distribution modified by the proton-$\eta$ interaction,
     with a scattering length $a_{p\eta}$~=~0.7~fm~+~$i~0.3$~fm.
     Details of the calculations together with the discussion of the nucleon-nucleon
     and nucleon-meson final-state-interaction can be found in reference~\cite{review}. 
  }
}
\end{figure}
 The interaction between particles depends on their relative momenta
 or equivalently on the invariant masses of the two-particle subsystems.
 Only two of the  invariant masses of the three subsystems are independent.
 Therefore the 
 entire principally accessible information about the final state
 interaction of the three--particle system can be presented in the form of the Dalitz plot.
Figure~\ref{dalitz_mc}(left) indicates 
the event distribution over the available surface in the phase-space
expected for the $pp \eta$ system at an excess energy of $\mbox{Q} = 16\,\mbox{MeV}$,
assuming a homogeneous primary production and taking into account the S--wave
interaction between the protons.
The proton--proton FSI modifies the homogeneous Dalitz plot distribution of
``non--interacting particles'', enhancing its population at a region where the
protons have small relative momenta.
Figure~\ref{dalitz_mc}(right) shows the phase-space density
distribution simulated when
switching off the proton--proton interaction but accounting for the interaction
between the $\eta$--meson and the proton.
Due to the lower strength of this interaction the expected deviations from a
uniform distribution is by about two orders of magnitude smaller, but still one
recognizes a slight enhancement of the density in the range of low invariant
masses of proton--$\eta$ subsystems.
However, due to weak variations of the proton--$\eta$ scattering amplitude the
enhancement originating from the $\eta$--meson interaction with one proton is
not separated from the $\eta$--meson interaction with the second proton.
Therefore an overlapping of broad structures occurs.
It is observed that the occupation density grows slowly with increasing
$\mbox{s}_{pp}$ opposite to the effects caused by the S--wave proton--proton
interaction, yet similar to the modifications expected for the P--wave
one~\cite{dyringPHD}.
From the above example it is obvious that only 
in experiments with a high statistics, 
signals from the meson--nucleon interaction can appear over the overwhelming
nucleon--nucleon final state interaction.
\section{Experimental results}
\begin{figure}[h]
\parbox{0.5\textwidth}{\centerline{
        \epsfig{figure=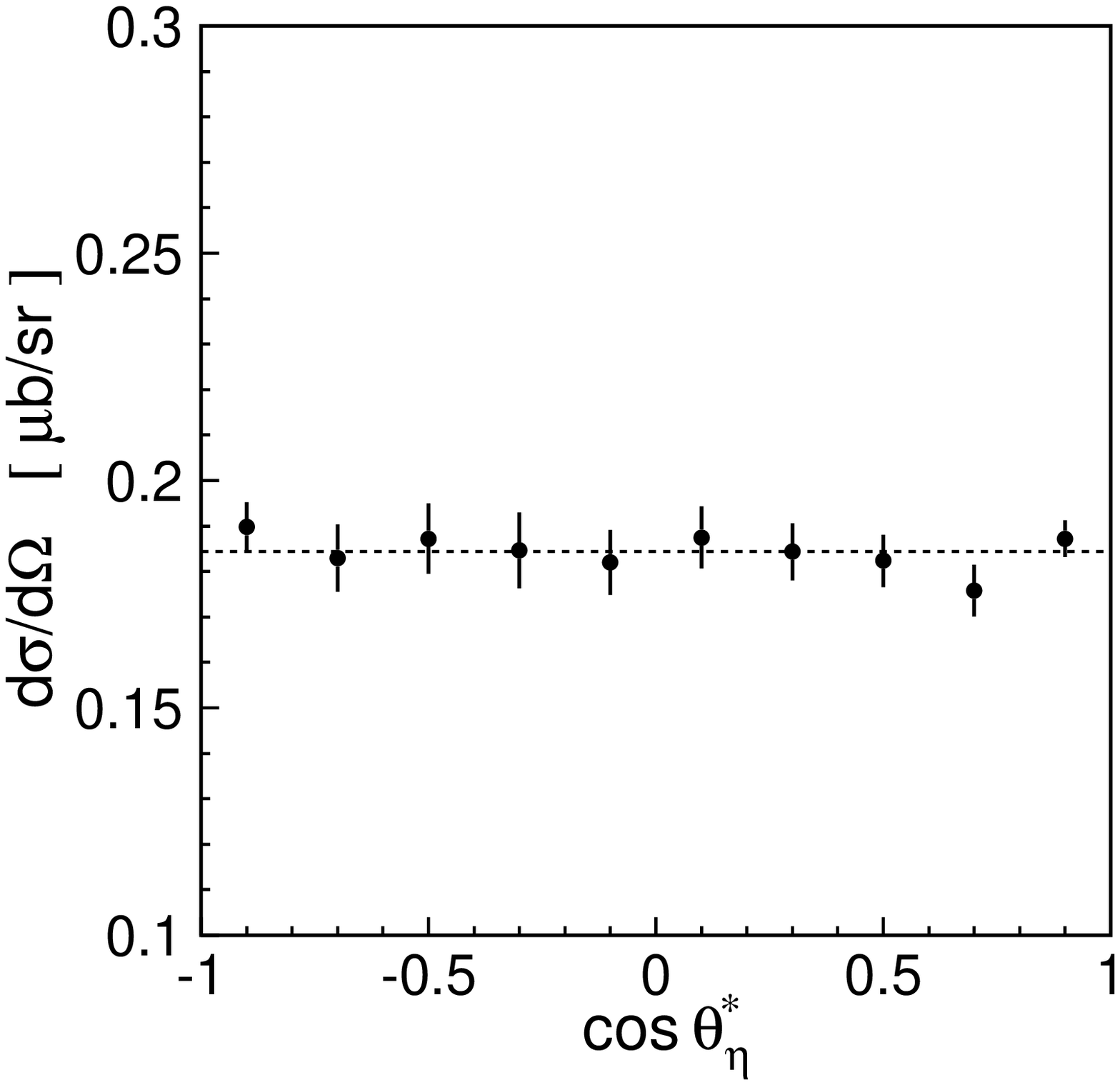,width=0.45\textwidth,angle=0}}
}
\parbox{0.5\textwidth}{\centerline{
        \epsfig{figure=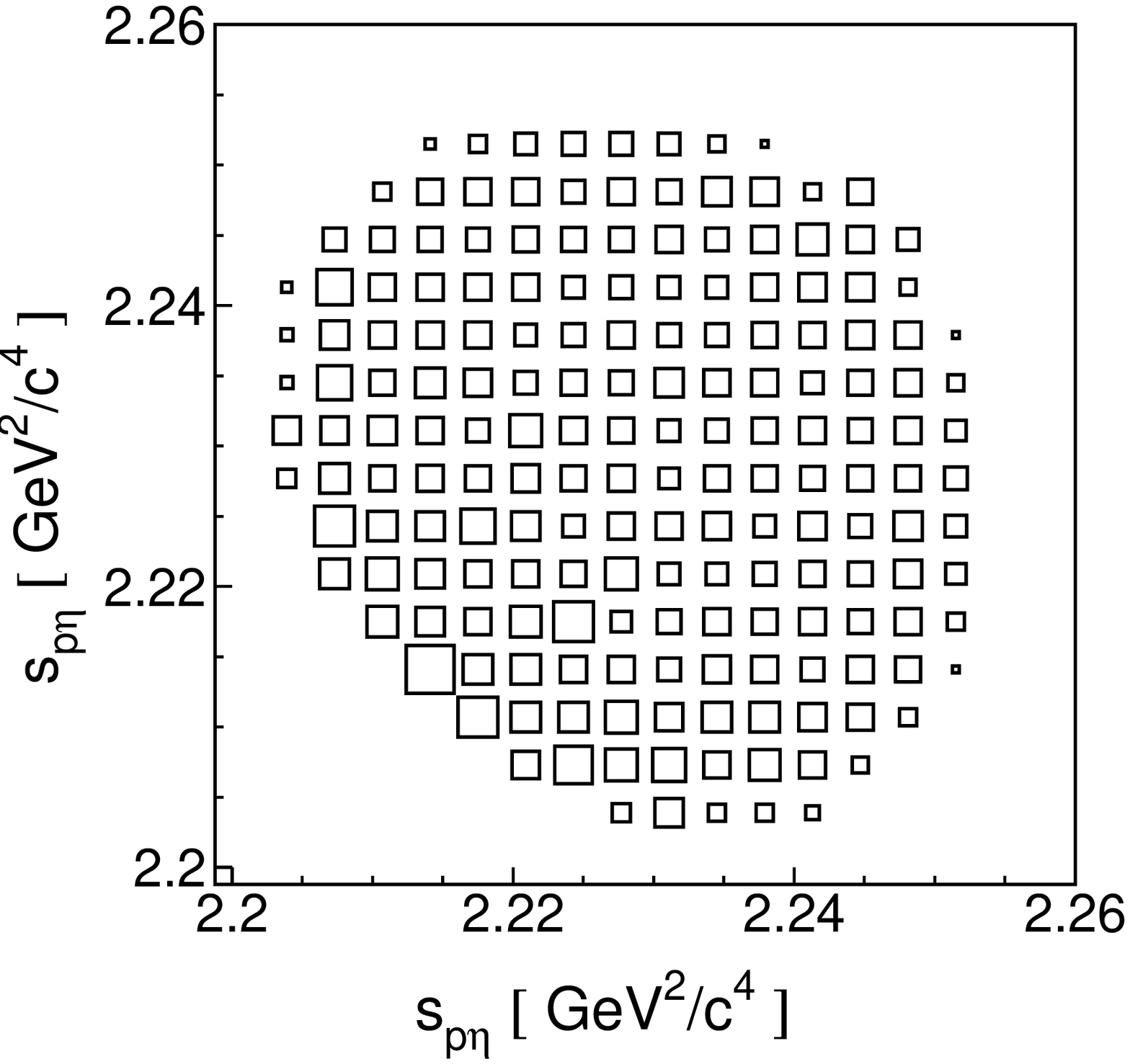,width=0.45\textwidth,angle=0}}
}
\parbox{1.0\textwidth}{
  
 \vspace{-0.2cm}
  \caption{ \label{dalitz_exp}
     (left) Experimentally determined 
      differential cross section of the $pp \to pp\eta$ reaction as a function 
     of the $\eta$ meson center-of-mass polar angle at Q~=~15.5~MeV.
     (right) Dalitz-plot distribution corrected for the detection acceptance
     and the proton-proton final-state-interaction. The proton-proton FSI
     enhancement factor has been calculated as the inverse of the Jost function 
     given in reference~\cite{goldberger}.
  }
}
\end{figure}

\begin{figure}[H]
\parbox{0.5\textwidth}{\centerline{
        \epsfig{figure=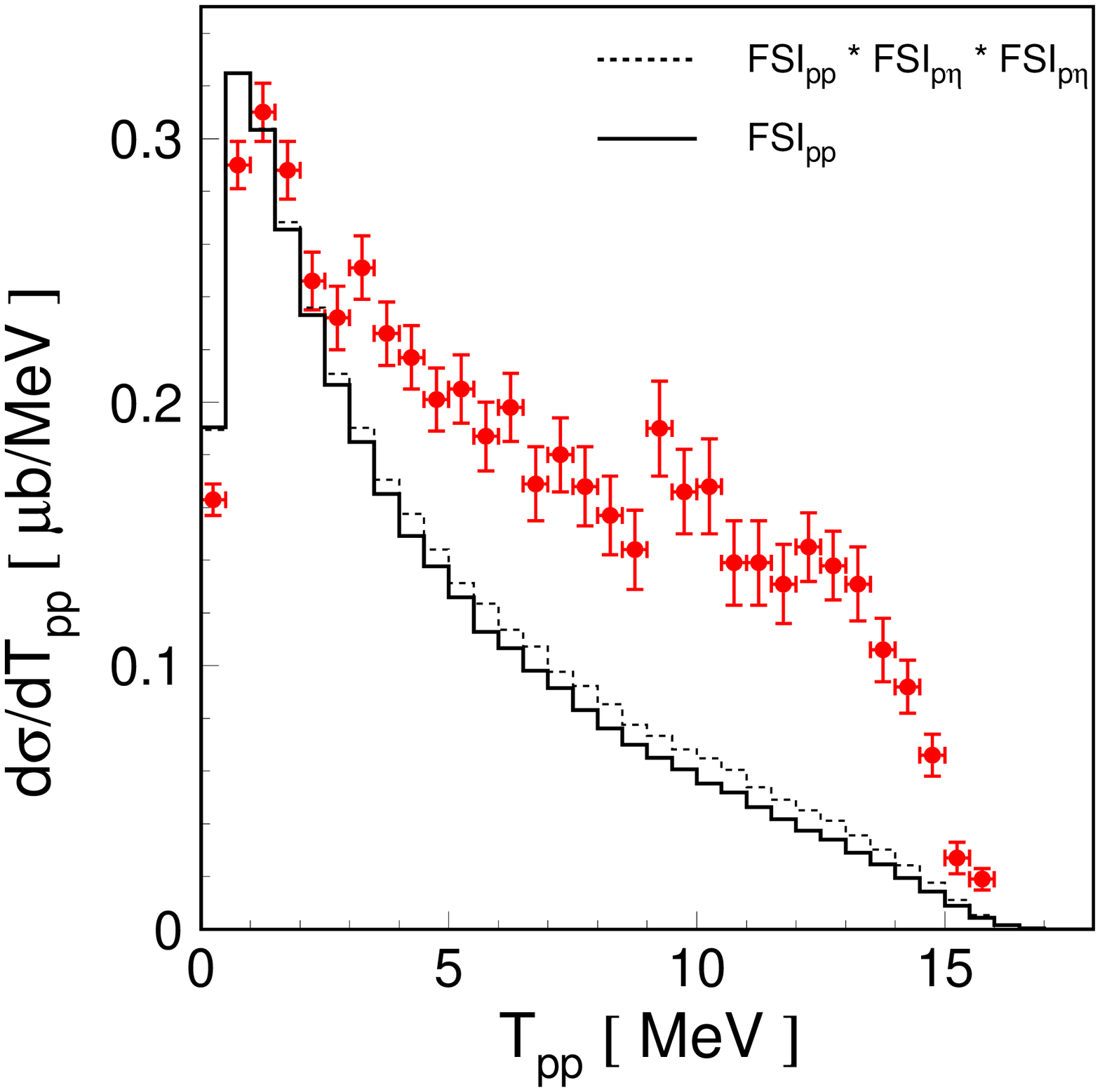,width=0.45\textwidth,angle=0}}
}
\parbox{0.5\textwidth}{\centerline{
        \epsfig{figure=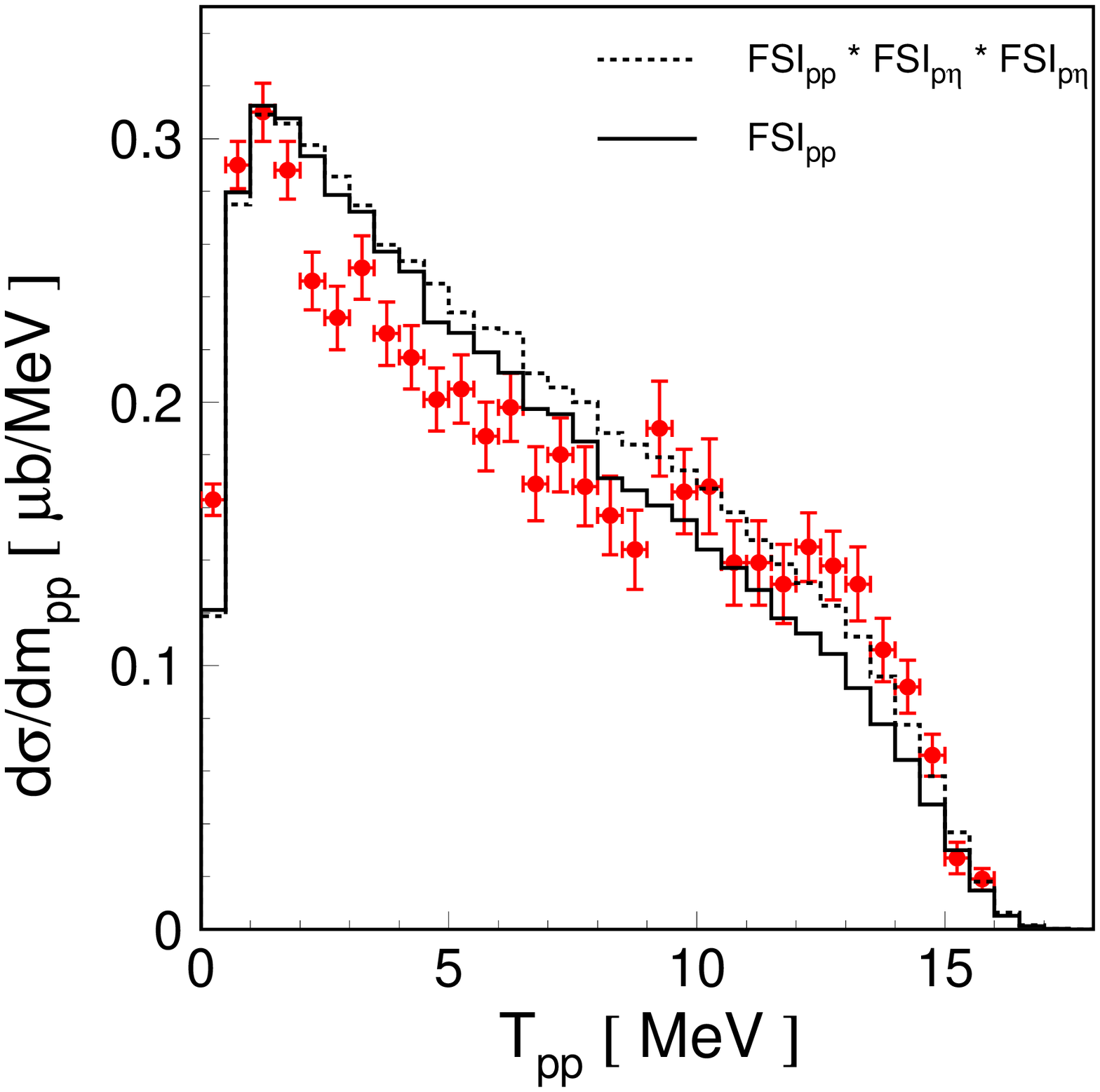,width=0.45\textwidth,angle=0}}
}
\parbox{1.0\textwidth}{
  
  \vspace{-0.2cm}
  \caption{ \label{dsigmapodT}
          Experimental distribution of the kinetic energy 
          ($T_{pp}~=~\sqrt{s_{pp}} - 2m_{p}$)
          in the proton-proton  subsystem determined experimetally 
          for the $pp  \to pp\eta$ reaction at the excess energy of Q~=~15.5~MeV.
          The lines represent calculations as described in the text.
          They have been roughly normalized 
          to the data at $T_{pp}$ below 3~MeV. 
  }
}
\end{figure}

\begin{figure}[H]
\parbox{0.333\textwidth}{
        \epsfig{figure=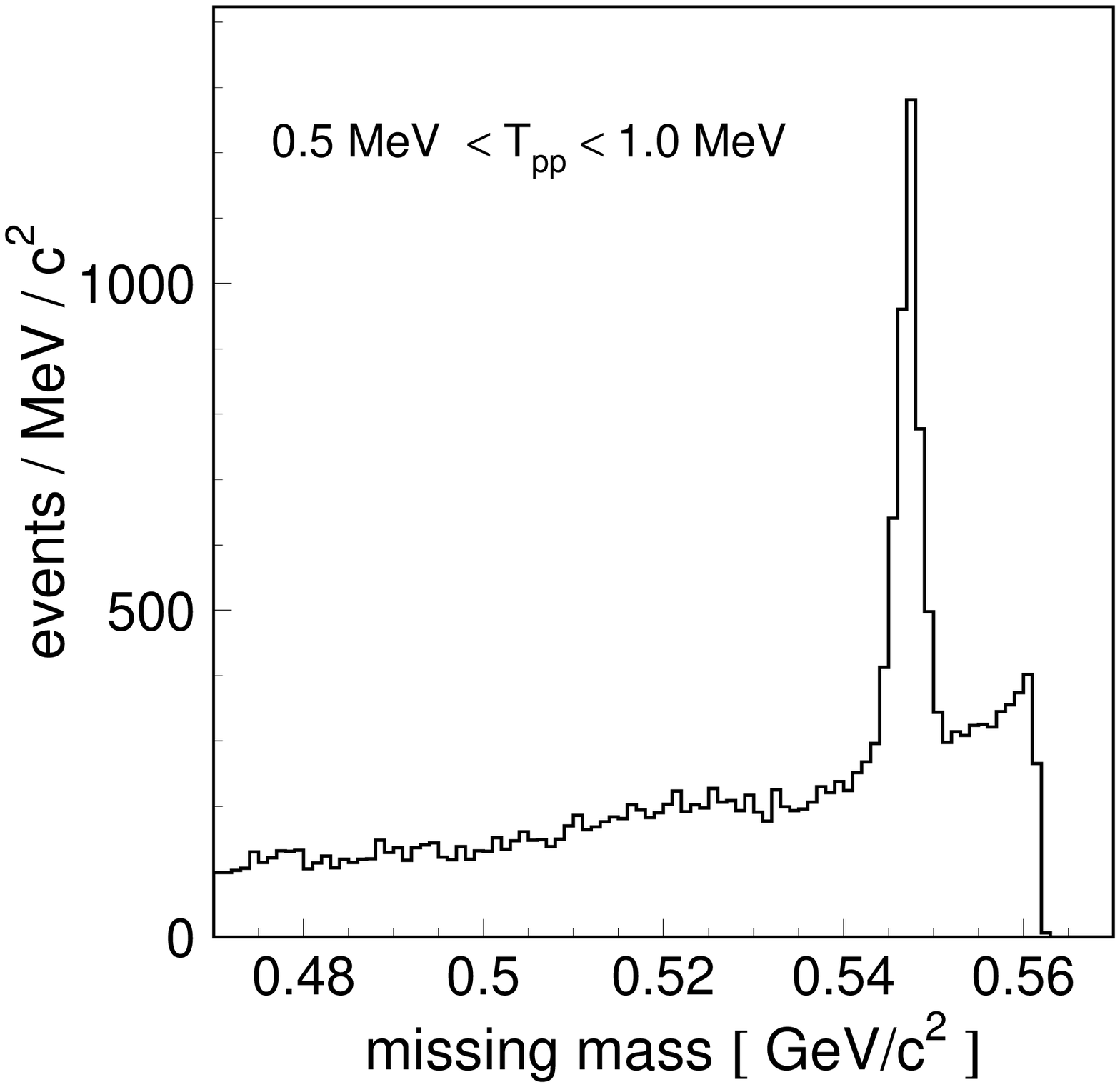,width=0.35\textwidth,angle=0}
}
\parbox{0.333\textwidth}{
        \epsfig{figure=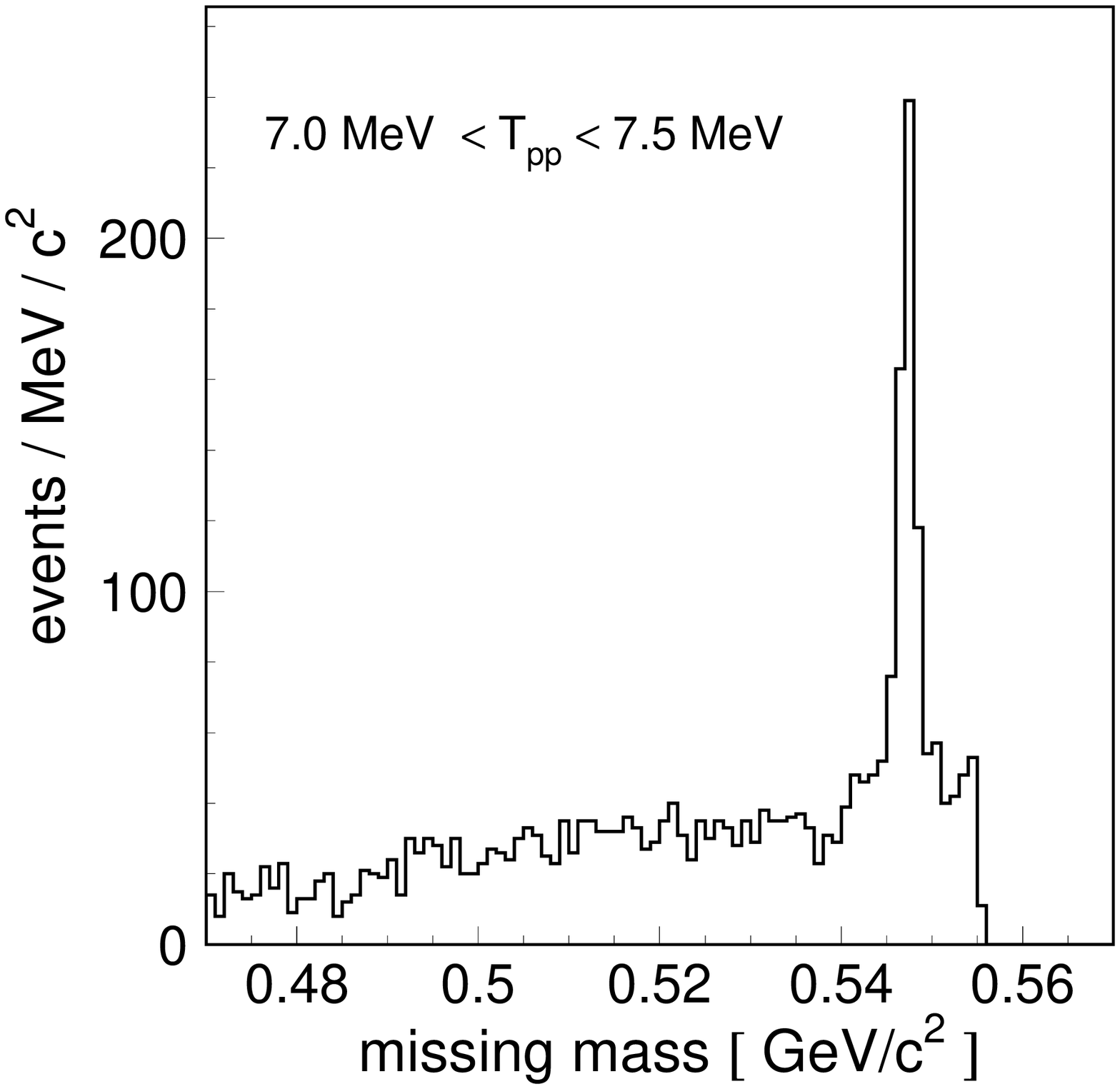,width=0.35\textwidth,angle=0}
}
\parbox{0.333\textwidth}{
        \epsfig{figure=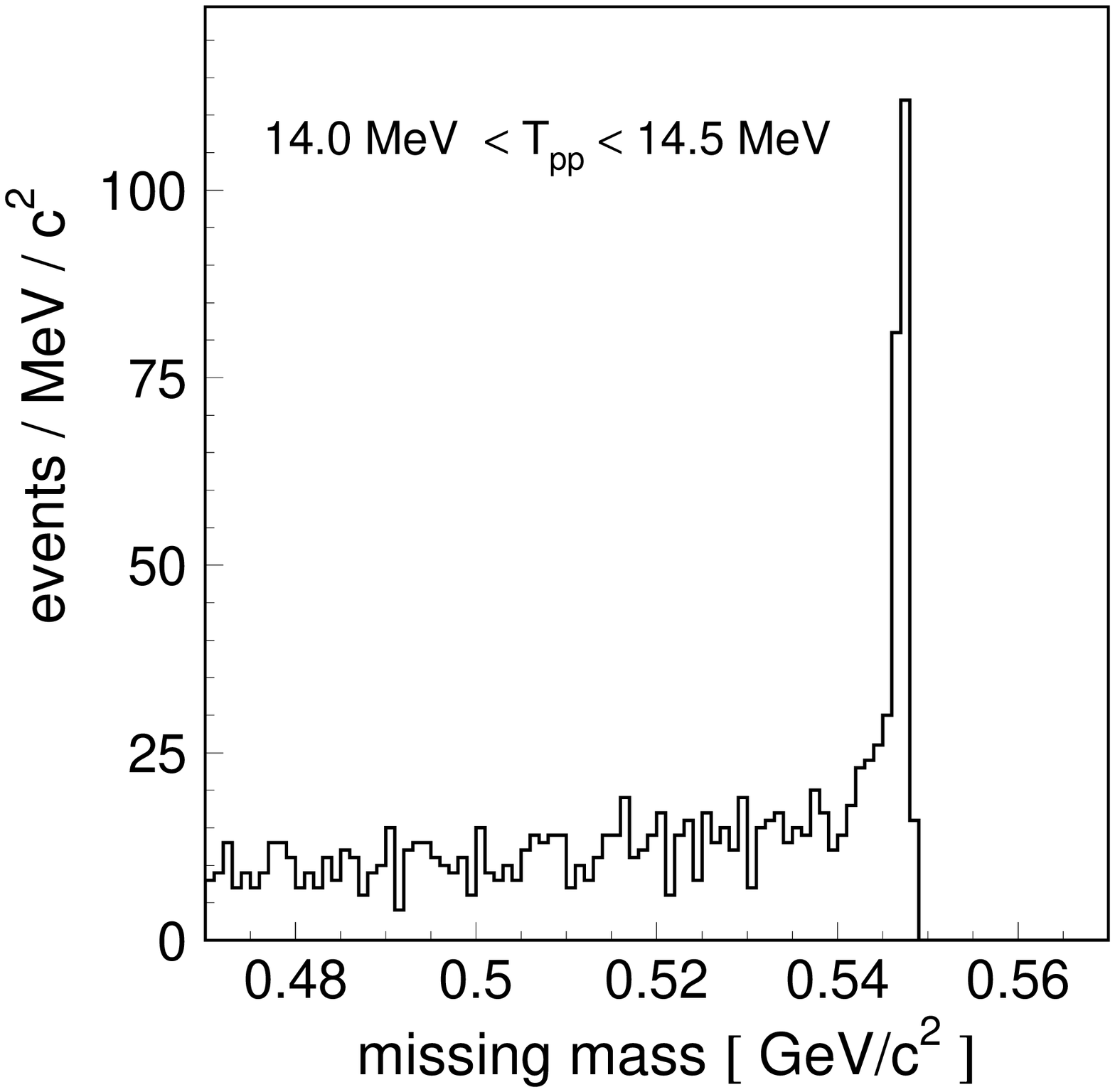,width=0.35\textwidth,angle=0}
}
\parbox{1.0\textwidth}{
  \caption{ \label{misski}
   Example of a missing mass distribution obtained for the $pp\to ppX$
   reaction at three $T_{pp}$ regions 
   for small, intermediate and large $T_{pp}$ values.  
   A sharp peak corresponds to the $pp\to pp\eta$ events. The broad distribution 
   is due to the multi-pion production via the 
   reactions $pp\to pp2\pi$ and $pp\to pp3\pi$. 
   Its contribution to the $\eta$-peak can not be discriminated 
   on the event-by-event basis by means of the missing mass technique.
  }
}
\end{figure}

 The enhancement observed
 in the total cross section encouraged us to perform 
 the high statistics measurements
 of the $pp \to pp\eta$ reaction in order to investigate a possible manifestation 
 of the $\eta$-nucleon-nucleon dynamics in the occupation of the available phase-space.
 Here we report on measurements  
 of the $pp \to pp\eta$ reaction at an excess energy 
 of Q~=~15.5~MeV.  The large number of identified $pp \to pp\eta$ events (24000)
 permits a statistically significant determination of the differential cross sections.
 The acceptance of the detection system covers 
 the full range of the $\eta$ meson center-of-mass polar scattering angles~\cite{eta_menu},
 and enables to prove that at this excess energy~\cite{review}  the $\eta$ meson
 is produced completely isotropically in the reaction center-of-mass 
 system (fig.~\ref{dalitz_exp}(left)),  as expected. 
 Figure~\ref{dalitz_exp}(right) presents the Dalitz
 plot of the identified $pp\eta$ system corrected for the detection acceptance and the
 proton-proton final-state-interaction. One recognizes 
 an increase of the occupation density at small
 values of $s_{p\eta}$. 
\vspace{-0.9cm}
 The observed effect is much stronger than 
 the one obtained from the simulations performed 
 under the assumption that the overall FSI effect can be 
 separated from the primary production and that the overall enhancement factor
 can be factorized into the incoherent pairwise interactions.
 A deviation of the experimentally observed  population of the phase-space 
 from the expectation based on the mentioned assumptions
 is even better visible in figure~\ref{dsigmapodT}. 
\vspace{-0.2cm}
 This figure presents the  projection 
 of the phase-space distribution onto the $T_{pp}~=~\sqrt{s_{pp}} - 2m_{p}$ axis corresponding
 to the axis indicated 
 by the arrows in the two parts of figure~\ref{dalitz_mc}. 
 The statistics allowed to identify 
 the number of $pp \to pp\eta$ events in bins of 0.5~MeV 
 of the kinetic energy of protons $T_{pp}$ in their rest frame. 
 As an example the missing mass spectra corresponding to the large, small and middle 
 values of $T_{pp}$ are shown in figure~\ref{misski}.
\vspace{-0.3cm}
 At each presented spectrum
 a signal originating from the
 $pp\to pp\eta$ reaction is evidently seen over a smooth distribution and allows for the
 model independent determination of the $d\sigma/dT_{pp}(T_{pp})$ distribution~(fig.~\ref{dsigmapodT}).
 The superimposed lines in figure~\ref{dsigmapodT} correspond to the calculations
 performed under the assumption that the production amplitude can be factorized
 into primary production and the final state interaction.
 The solid lines depict calculations where only the proton-proton FSI was taken into account,
 whereas the dashed lines present results where the overall enhancement
 was factorized into the corresponding pair interactions of the $pp\eta$ system.
 In the left panel the enhancement factor accounting 
 for the proton-proton FSI has been  calculated
 as a square of the on-shell proton-proton scattering amplitude
 derived according to the modified Cini-Fubini-Stanghellini  formula including
 Wong-Noyes Coulomb corrections~\cite{noyes995,swave,review}, whereas in the right panel
 the inverse of the Jost function presented in references~\cite{goldberger,niskanen107}
 was used. 
 Though the simple phenomenological treatment --~based on the factorization of the
 production amplitude into the constant primary production and the on-shell
 incoherent pairwise interaction among the exit particles~-- works astonishingly well 
 in case of the total cross section energy dependence~\cite{review},
 it fails completely in the description of the differential cross section 
 as can be inferred from figure~\ref{dsigmapodT}(left).  Taking instead the 
 inverse of the Jost function as an
 enhancement factor of the proton-proton interaction,
  which should  account approximately for the off-shell effects,
 one obtaines a much better agreement with the data. However, 
 the experimentally determined structure
 is not satisfyingly reproduced, 
 and calls for a more sophisticated theoretical 
 interpretation. The preliminary theoretical study 
 indicates that the effect is selective for the 
 the primary  production 
 mechanism~\cite{kanzo}.
 It is worth to note that the obtained results are in 
 agreement with the observation performed by the TOF collaboration
 using a completely different detection techinque~\cite{TOFeta}.
\vspace{-0.4cm}
\section{Acknowledgement}
{\small
 The work has been partly supported by the European Community - Access to
Research Infrastructure action of the Improving Human Potential Programme 
}

\end{document}